\documentclass[10pt]{article}

\usepackage{a4,amssymb,amsmath,graphicx,wrapfig}
\usepackage{vmargin}
\setpapersize{custom}{210mm}{297mm}
\newcommand{\z}[1]{{#1}}
\newcommand{\RRR}{\mathbb{R}}

\newcommand{\NNN}{\mathbb{N}}
\newcommand{\CCC}{\mathbb{C}}

\newcommand{\bx}{\boldsymbol x}

\newcommand{\bxp}{\boldsymbol x'}

\newcommand{\bg}{\boldsymbol g}

\newcommand{\bk}{\boldsymbol k}
\newcommand{\bkt}{\boldsymbol {\widetilde{k}}}
\newcommand{\kt}{\widetilde{k}}
\newcommand{\bks}{\boldsymbol k_{\text{s}}}
\newcommand{\ks}{k_{\text{s}}}

\newcommand{\by}{\boldsymbol y}
\newcommand{\be}{\boldsymbol e}

\newcommand{\bxi}{\boldsymbol \xi}

\newcommand{\bj}{\boldsymbol j}
\newcommand{\bsig}{\boldsymbol \sigma}

\newcommand{\bn}{\boldsymbol n}
\newcommand{\bom}{\boldsymbol \omega}

\newcommand{\kstatb}{k_{\text{stat}}}

\newcommand{\pin}{\psi_{\text{in}}}
\newcommand{\pout}{\psi_{\text{out}}}
\newcommand{\cout}{\chi_{\text{out}}}
\newcommand{\pouth}{\widehat{\psi}_{\text{out}}}
\newcommand{\couth}{\widehat{\chi}_{\text{out}}}
\newcommand{\phip}{\varphi_{+}}
\newcommand{\ran}{\text{Ran}}
\newcommand{\dom}{\text{D}}

\newcommand{\kfac}{\kappa}
\newcommand{\kks}{k-\be_{\bk}\cdot\bks}
\newcommand{\gprime}{\mathcal{G}^0}

\newcommand{\ghut}{\mathcal{G}^{+}}
\newcommand{\ltr}{L^2(\RRR^3)}
\newcommand{\lxr}{\langle x\rangle}
\newcommand{\lkr}{\langle k\rangle}

\newcommand{\supp}{\operatorname{supp}}

\newcommand{\limmean}{\operatorname{l.i.m.}}

\newsavebox{\proofnumber}

\newenvironment{proof}[1]{\vspace{0,5cm}\parindent=0pt\sbox{\proofnumber}{\sl #1}{\sl Proof}\usebox{\proofnumber}.\parindent=15pt }{\vspace{0,5cm}}

\newtheorem{theorem}{Theorem}

\newcommand{\bthe}{\begin{theorem}\hspace{-1.1ex}{\bf .}\hspace{2ex}}
\newcommand{\ethe}{\end{theorem}}

\newcommand{\bpro}{\begin{proof}}

\newcommand{\epro}{$\blacksquare$\end{proof}}

\newtheorem{lemma}{Lemma}

\newcommand{\blem}{\begin{lemma}\hspace{-1.1ex}{\bf .}\hspace{2ex}}
\newcommand{\elem}{\end{lemma}}

\newtheorem{remark}{\sl Remark}

\newcommand{\brem}{\begin{remark}\rm\hspace{-1.1ex}{.}\hspace{2ex}}
\newcommand{\erem}{\end{remark}}

\newtheorem{definition}{Definition}

\newcommand{\bde}{\begin{definition}\hspace{-1.1ex}{\bf .}\hspace{2ex}}
\newcommand{\ede}{\end{definition}}

\newtheorem{corollary}{Corollary}

\newcommand{\bcor}{\begin{corollary}\hspace{-1.1ex}{\bf .}\hspace{2ex}}
\newcommand{\ecor}{\end{corollary}}

\newtheorem{proposition}{Proposition}

\newcommand{\bprop}{\begin{proposition}\hspace{-1.1ex}{\bf .}\hspace{2ex}}
\newcommand{\eprop}{\end{proposition}}

\newcounter{cond}

\newenvironment{cond}[0]{\refstepcounter{cond}
\begin{itemize}\item[\bfseries{A\arabic{cond}.}]}{\end{itemize}}

\newcommand{\bcond}{\begin{cond}}
\newcommand{\econd}{\end{cond}}

\begin{document}
\parindent=0pt
\thispagestyle{empty}

\vspace{1cm}

{\huge{\bfseries{The Flux-Across-Surfaces Theorem }}}\vspace{0,3cm}

{\huge{\bfseries{under conditions on the scattering}}}\vspace{0,3cm}

{\huge{\bfseries{state}}}\vspace{0,7cm}

D. D\"urr$^1$, T. Moser$^1$, P. Pickl$^2$

\begin{list}{1}{\setlength{\labelwidth}{2em}\setlength{\leftmargin}{1em}}
\item[1]Mathematisches Institut der Universit\"at M\"unchen,
\\Theresienstr. 39, 80333 M\"unchen, Germany\\ e-mail: duerr@mathematik.uni-muenchen.de, moser@mathematik.uni-muenchen.de\item[2]Mathematisches Institut der Universit\"at T\"ubingen,
\\Auf der Morgenstelle 10, 72076 T\"ubingen, Germany\\ e-mail: peter.pickl@uni-tuebingen.de
\end{list}

\today \vspace{1cm}

{\bfseries{\small{Abstract. }}}{\small{The flux-across-surfaces
theorem (FAST) describes the outgoing asymptotics of the quantum flux density of a scattering state. The FAST has been proven for potential scattering under conditions on
the outgoing asymptote $\pout$ (and of course under suitable conditions on the scattering potential).  In this article we prove
the FAST under conditions on the scattering state itself. \z{In the proof we will establish also new mapping properties of the wave operators}.}}

\parindent=15pt
\section{Introduction}\label{chintro}

The flux-across-surfaces theorem (FAST) is  basic to the empirical
content of scattering theory. The FAST describes the relation
between the integrated quantum flux density of a scattering state over a
(detector) surface and a (detection) time interval and the momentum
distribution of the corresponding outgoing asymptote $\pout$. In this paper we deal with
the simplest case of one-particle potential \z{scattering}. We remark that the extension of the FAST to many-particle scattering theory is problematical, see \cite{duerrteufel:02}.

With the
quantum flux density ($^*$ denotes the complex conjugate)
$$\bj^{\psi}=\operatorname{Im}(\psi^*\nabla\psi)$$ and without spelling out the conditions under which it can be proven, the FAST reads 
\begin{equation}\label{fluxi}
  \lim\limits_{R\to\infty}\int\limits_T^{\infty}\int\limits_{R\Sigma}\bj^{\psi}(\bx,t)
  \cdot d\bsig dt=
  \lim\limits_{R\to\infty}\int\limits_T^{\infty}\int\limits_{R\Sigma}\left |\bj^{\psi}(\bx,t)
  \cdot d\bsig\right| dt
  =\int\limits_{C_{\Sigma}}|\pouth(\bk)|^2d^3k,
\end{equation}
where $\Sigma\subset S^2$ is a subset of the unit sphere, $R\Sigma:=\{\bx\in\RRR^3:\bx=R\bom,\;\bom\in\Sigma\}$ is the spherical surface covering the solid angle $\Sigma$ and $C_{\Sigma}:=\{\bk\in\RRR^3:\be_k\in\Sigma\}$ is the cone
given by $\Sigma.$ Furthermore $\hat{ }$ denotes the Fourier transform and $\pout$ the outgoing asymptote to the corresponding scattering state $\psi=\Omega_+\pout$ with the wave operator $\Omega_+$. 

The left hand side is interpreted and also shown to be the crossing probability of the
 particle crossing the surface $R\Sigma$ \cite{cohen2:97,combes:75,newton:82}, \cite{berndl:94,daumer:96,duerr:00}. From the
 crossing probability one derives the scattering cross section \cite{duerr2:04,duerr:00}. The right hand side of (\ref{fluxi}) relates the
  crossing probability to the $S$-matrix. Technically, the FAST (\ref{fluxi}) has been proven requiring conditions on $\pout$. But it is clear that when all is said and done one needs the conditions on the scattering state for which the FAST holds. In particular, the microscopic derivation
  of the cross section needs the FAST under conditions on the scattering state \cite{duerr2:04}. In the
  present paper we establish the FAST (\ref{fluxi}) under conditions on the scattering state.

\z{The} FAST has been put into a mathematically rigorous setting by
Combes, Newton, Shtokhammer in 1975 \cite{combes:75}. In 1996 the FAST was
proven by Daumer et al. \cite{daumer:96} for the Schr\"odinger case
without a potential. One year later Amrein, Pearson and Zuleta
proved the FAST for short and long range potentials using methods in
the context of Kato's $H$-smoothness, requiring an energy
cutoff on the outgoing asymptote \cite{amrein1:97,
amrein2:97}. (More precisely, $\supp\pouth$ is compact.) This at first sight innocently looking requirement seems however to be an important hindrance towards the physically relevant
formulation of the FAST with conditions on the scattering state
itself. We shall discuss this in somewhat more detail later. In 1999 Teufel, D\"urr and Berndl gave a proof based on
eigenfunction expansions \z{without an energy cutoff, but still requiring} smoothness properties of the outgoing asymptote for potentials falling off faster than $x^{-4}$ \cite{teufel2:99}. Panati and Teta gave a proof for the special
case of point interactions under conditions on the scattering state
\cite{panati:00} with similar methods as in \cite{teufel2:99}. In 2003 Nagao \cite{nagao:04} proved a
 weaker result, namely leaving out  the second equality in
equation (\ref{fluxi}). This proof works for short range potentials falling off faster than the dimension of the space ($=3$) and requires only conditions on the scattering state. By leaving out the second equality in (\ref{fluxi}) the result does not establish the connection to empirical data of a typical scattering experiment, as it does not establish the probabilistic meaning of
the quantum flux as a crossing probability or in technical terms it
does not establish that the flux points asymptotically outwards.  In
the same year D\"urr and Pickl \cite{duerr:03} proved the FAST for a Dirac particle under conditions on the scattering state alone using eigenfunction expansions.
 

We provide now a proof for the Schr\"odinger case combining the techniques of the proofs in \cite{duerr:03,teufel2:99}
  to establish the FAST under conditions on the scattering state and for potentials falling off faster than $x^{-4}.$ The idea is to prove the FAST under almost optimal conditions on $\pout$,
  which can be translated to reasonable and easily checkable conditions on the scattering state.
It is clearly essential that there is no energy cutoff on $\pout$,
because it is highly unclear whether there are any reasonable
conditions on the scattering state ensuring a cutoff on $\pout$ (cf. (\ref{mapi})). Having formulated the task to prove the FAST
 under conditions on the outgoing asymptote which can be transferred to conditions on the scattering
 state we like to remark,
that there are no suitable assertions in the literature which allow
to transfer conditions on $\psi$ to $\pout$ in the context of the
proof of the FAST.\footnote{For mapping properties between $\psi$ and $\pout$, which are not applicable in our case, see e.g. \cite{yajima:95}.} We shall elaborate this further considering
eigenfunction expansions. We recall the generalized Fourier
transform (see Lemma \ref{lemexpansion}), which maps the scattering
state $\psi$ to the ordinary Fourier transform $\pouth$ of $\pout$ :
\begin{equation}\label{mapi}
\pouth(\bk)=(2\pi)^{-\frac{3}{2}}\int\varphi^*_+(\bx,\bk)\psi(\bx)d^3x,
\end{equation}
where $\varphi^*_+(\bx,\bk)$ are the generalized eigenfunctions.
In Lemma \ref{lemlippschw} we collect the properties of the
eigenfunctions which are---concerning smoothness and
boundedness---in general very poor: The generalized eigenfunctions
are solutions of the Lippmann-Schwinger equations:
\begin{equation}\label{lippmanni}
\varphi_{\pm}(\bx,\bk)=e^{i\bk\cdot\bx}-\frac{1}{2\pi}\int\frac{e^{\mp
  ik|\bx-\bxp|}}{|\bx-\bxp|}V(\bxp)\varphi_{\pm}(\bxp,\bk)d^3x',
\end{equation}
in which we note the appearance of the absolute value $k$ of $\bk$
in the spherical wave part. Derivatives of $k$ of higher
order than one behave singular for $k\to 0.$ Therefore we expect in general that the derivatives of the generalized eigenfunctions (of higher order than one) are unbounded for small $k$.\footnote{For a point interaction the generalized eigenfunctions can be explicitly computed \cite{albeverio:88}, p. 39. They have exactly this singular behavior.}
In view of (\ref{mapi}) this  singular behavior is typically
inherited by $\pout$ and it is hard to see how ``extreme'' conditions on $\pout$ like
$\pout$ in Schwartz space or $\pouth$ compactly supported can be satisfied by reasonable scattering potentials or states. This caveat applies to
the above cited works on the FAST except \cite{duerr:03,nagao:04,panati:00}. Our task is thus to read from (\ref{mapi}) proper conditions  on
$\pout$ which can be formulated in terms of the scattering state
and then to prove the FAST under these conditions. 

The paper is organized as follows: In Section \ref{chsetup} we
recall the mathematical basics of scattering theory including recent results and fix
notations, in Section \ref{chflux}, we formulate and prove the FAST
under \z{weaker} conditions on the asymptote than in
\cite{teufel2:99}. The conditions will be transformed by the
mapping Lemma \ref{lemmapping} to sufficient conditions on the
scattering state. The most complete statement is Corollary \ref{corflux}.
 Technically the FAST is proven by stationary phase
methods, which turns out---due to our necessarily weak
conditions---to be a rather involved modification of standard
results, e.g. Theorem 7.7.5 in \cite{hoermander:83}. The proof of
the modified assertion is done in the appendix.

\section{The mathematical framework of potential scattering}\label{chsetup}
\sectionmark{The mathematical framework}
We list those results of scattering theory (e.g. \cite{amrein:77,
duerr:01,ikebe:60, kato:51, pearson:88, simon3:79, simon1:80,
teufel2:99}) which are essential for the proof of the FAST in Section \ref{chflux}.

We use the usual description of a nonrelativistic spinless system
by the Hamiltonian $H$\label{hamiltonian} (we use natural units
$\hbar=m=1$):
\begin{equation*}
  H:=-\frac{1}{2}\Delta+V(\bx)=:H_0+V(\bx),
\end{equation*}
with the real-valued potential\label{potential} $V\in (V)_n$, defined as follows:
\bde\label{defpot} $V$ is in $(V)_n$, n=2,3,4,..., if
\begin{itemize}
\item[(i)] $V\in \ltr$,\item[(ii)] $V$ is locally H\"older continuous except at a finite
number of singularities, 
\item[(iii)] there exist positive numbers
$\epsilon,\;C_0,\;R_0$ such that
\begin{equation*}
  |V(\bx)|\leq C_0\lxr^{-n-\epsilon}\text{ for }|\bx|\geq R_0,
\end{equation*}
where $\langle\cdot\rangle:=(1+(\cdot)^2)^{\frac{1}{2}}.$
\end{itemize}
\ede
Under these conditions (see e.g. \cite{kato:51})  $H$ is
self-adjoint on the domain D($H)=$D($H_0)=\{f\in
\ltr:\int|k^2\widehat{f}(\bk)|^2d^3k<\infty\}$, where
$\widehat{f}:=\mathcal{F}f$ is the Fourier transform:
\begin{equation}\label{fourier}
  \widehat{f}(\bk):=(2\pi)^{-\frac{3}{2}}\int
e^{-i\bk\cdot\bx}f(\bx)d^3x.
\end{equation}
Let $U(t)=e^{-iHt}$. Since $H$ is self-adjoint on the domain
D($H$), $U(t)$ is a strongly continuous one-parameter unitary
group on $\ltr$. Let $\phi\in$D($H$). Then $\phi_t\equiv
U(t)\phi\in$D($H$) and satisfies the Schr\"odinger equation:
\begin{equation*}
  i\frac{\partial}{\partial   t}\phi_t(\bx)=H\phi_t.
\end{equation*}
We define the wave operators $\Omega_{\pm}$ with the range $\ran(\Omega_{\pm})$ in the usual way:
\begin{align*}
\Omega_{\pm}&:\ltr\to\ran(\Omega_{\pm}),\\
 \Omega_{\pm}&:=\text{s-}\lim\limits_{t\to\pm\infty}e^{iHt}e^{-iH_0t},
\end{align*}
where $\text{s-}\lim$ denotes the limit in the $L^2$-sense.
Ikebe \cite{ikebe:60} proved that for a potential $V\in (V)_2$ the
wave operators exist and have the range (this property is called asymptotic completeness):
\begin{equation*}
\text{Ran}(\Omega_{\pm})=\mathcal{H}_{\text{cont}}(H)=\mathcal{H}_{\text{a.c.}}(H),
\end{equation*}
where $\mathcal{H}_{\text{cont}}(H)$ and $\mathcal{H}_{\text{a.c.}}(H)$ denote the subspaces of $\ltr$ consisting of states, which belong to the continuous and the absolutely continuous part of the spectrum of $H.$
Then we have for every $\psi\in\mathcal{H}_{\text{a.c.}}(H)$ asymptotes $\pin,\pout\in\ltr$ with:
\begin{equation}\label{defout}
  \Omega_-\psi_{\text{in}}=\psi=\Omega_+\psi_{\text{out}}.
\end{equation}
On D($H_0$) the wave operators satisfy  the so called intertwining
property
\begin{equation*}
 H\Omega_{\pm}=\Omega_{\pm}H_0.
\end{equation*}
On $\mathcal{H}_{\text{a.c.}}(H)\cap$D($H$) we have then that
\begin{equation}\label{inter}
 H_0\Omega_{\pm}^{-1}=\Omega_{\pm}^{-1}H.
\end{equation}
We will need the time evolution of a state
$\psi\in\mathcal{H}_{\text{a.c.}}(H)$ with the Hamiltonian $H$.
Its diagonalization on $\mathcal{H}_{\text{a.c.}}(H)$ is given by
the eigenfunctions $\varphi_{\pm}$\label{lipp}:
\begin{equation}\label{schrlipp}
(-\frac{1}{2}\Delta+V(\bx))\varphi_{\pm}(\bx,\bk)=\frac{k^2}{2}\varphi_{\pm}(\bx,\bk).\end{equation}
Applying $(-\frac{1}{2}\Delta-\frac{k^2}{2}\mp i0)^{-1}$ in (\ref{schrlipp}) one obtains the
Lippmann-Schwinger equation.
We recall the main parts of a result on this due to Ikebe in
\cite{ikebe:60} which is collected in the present form in
\cite{teufel2:99}.

\blem\label{lemexpansion} Let $V\in (V)_2$. Then for any
$\bk\in\RRR^3\backslash\{0\}$ there are unique solutions
$\varphi_{\pm}(\cdot,\bk):\RRR^3\to\mathbb{C}$ of the
Lippmann-Schwinger equations
\begin{equation}\label{lippmann}
\varphi_{\pm}(\bx,\bk)=e^{i\bk\cdot\bx}-\frac{1}{2\pi}\int\frac{e^{\mp
  ik|\bx-\bxp|}}{|\bx-\bxp|}V(\bxp)\varphi_{\pm}(\bxp,\bk)d^3x',
\end{equation}
with the boundary conditions
$\lim_{|\bx|\to\infty}(\varphi_{\pm}(\bx,\bk)-e^{i\bk\cdot\bx})=0$,
which are also classical solutions of the stationary Schr\"odinger
equation (\ref{schrlipp}), such that:
\begin{itemize}
\item[(i)] For any $f\in \ltr$ the generalized Fourier
transforms\label{general}\footnote{$\limmean\int$ denotes $\text{s-}\lim\limits_{R\to\infty}\int\limits_{x\leq R}.$}
\begin{equation*}
  (\mathcal{F}_{\pm}f)(\bk)=\frac{1}{(2\pi)^{\frac{3}{2}}}\limmean
  \int\varphi_{\pm}^{\ast}(\bx,\bk)f(\bx)d^3x
\end{equation*}
exist in $\ltr$.
\item[(ii)] Ran($\mathcal{F}_{\pm})=\ltr$ and
$\mathcal{F}_{\pm}: \mathcal{H}_{\text{a.c.}}(H) \to \ltr $
are unitary and the inverse of $\mathcal{F}_{\pm}$ is given by
\begin{equation*}
(\mathcal{F}_{\pm}^{-1}f)(\bx)=\frac{1}{(2\pi)^{\frac{3}{2}}}\limmean
  \int\varphi_{\pm}(\bx,\bk)f(\bk)d^3k.
\end{equation*}
\item[(iii)] For any $f\in \ltr$ the relation
$\Omega_{\pm}f=\mathcal{F}_{\pm}^{-1}\mathcal{F}f$ hold, where
$\mathcal{F}$ is the ordinary Fourier transform given by
(\ref{fourier}).
\item[(iv)] For any
$f\in\operatorname{D}(H)\cap\mathcal{H}_{\text{a.c.}}(H)$ we
have:$$Hf(\bx)=\left(\mathcal{F}_{\pm}^{-1}\frac{k^2}{2}\mathcal{F}_{\pm}f\right)(\bx),$$
and therefore for any $f\in\mathcal{H}_{\text{a.c.}}(H)$
$$e^{-iHt}f(\bx)=\left(\mathcal{F}_{\pm}^{-1}e^{-i\frac{k^2}{2}t}\mathcal{F}_{\pm}f\right)(\bx).$$
\end{itemize}
\elem

In order to apply stationary phase methods we will need estimates
on the derivatives of the generalized eigenfunctions:

\blem\label{lemlippschw} Let the potential satisfy the condition
$(V)_n$ for some $n\geq 3.$ Then:
\begin{itemize}
\item[(i)]$\varphi_{\pm}(\bx,\cdot)\in
C^{n-2}(\RRR^3\setminus\{0\})$ for all $\bx\in\RRR^3$ and the
partial derivatives\footnote{We use the usual multi-index
notation:
$\alpha=(\alpha_1,\alpha_2,\alpha_3),\;\alpha_i\in\NNN_0,\;|\alpha|:=\alpha_1+\alpha_2+\alpha_3$\text{ and }$\partial_{\bk}^{\alpha}f(\bk):=
\partial_{k_1}^{\alpha_1}\partial_{k_2}^{\alpha_2}\partial_{k_3}^{\alpha_3}f(\bk).$}
$\partial_{\bk}^\alpha\varphi_{\pm}(\bx,\bk),$\\$|\alpha|\leq n-2,$
are continuous with respect to $\bx$ and $\bk.$
\end{itemize}
If, in addition, zero is neither an eigenvalue nor a resonance of
$H$, then
\begin{itemize}
\item[(ii)]$\sup\limits_{\bx\in\mathbb{R}^3,\bk\in\mathbb{R}^3}|\varphi_{\pm}(\bx,\bk)|<\infty$
\end{itemize}
and for any $\alpha$ with $|\alpha|\leq n-2$ there is a $c_{\alpha}<\infty$ such that
\begin{itemize}
\item[(iii)]$\sup\limits_{\bk\in\ \mathbb{R}^3\setminus\{0\}}
|\kfac^{|\alpha|-1}\partial_{\bk}^\alpha\varphi_{\pm}(\bx,\bk)| <
c_\alpha\lxr^{|\alpha|},\;\text{ with }\kfac:=\frac{k}{\lkr}.$
\end{itemize}
Similarly, for any $l\in \{1,...,n-2\}$ there is a $c_{l}<\infty$ such that
\begin{itemize}
\item[(iv)]$\sup\limits_{\bk\in\ \mathbb{R}^3\setminus\{0\}}
\left|\frac{\partial^l}{\partial k^l}\varphi_{\pm}(\bx,\bk)\right|
< c_l\lxr^{l}.$
\end{itemize}
\elem

\brem Zero is a resonance of $H$ if there exists a solution $f$ of $Hf=0$ such that $\lxr^{-\gamma}f\in\ltr$ for any $\gamma>\frac{1}{2}$ but not for $\gamma=0.$\footnote{There are various definitions, see e.g. \cite{yajima:95}, p. 552, \cite{albeverio:88}, p.20 and \cite{jensen:79}, p. 584.} The appearance of a zero eigenvalue or resonance can be regarded as an exceptional event: For a Hamiltonian $H=H_0+cV,\;c\in\RRR,$ this can only happen in a discrete subset of $\RRR,$ see \cite{albeverio:88}, p. 20 and \cite{jensen:79}, p. 589.\erem

\brem Lemma \ref{lemlippschw}, except the assertion (iii) was proved in
\cite{teufel2:99}, Theorem 3.1. Assertion (iii) repairs a false
statement in Theorem 3.1 which did not include the necessary
$\kfac^{|\alpha|-1}$ factor, which we have in (iii). For
$|\alpha|=1$ which was the important case in that paper there is
however no difference. For completeness we comment on the  proof
of this corrected version  in the appendix. We note that the problem which we address here comes from the appearance of the absolute value of $\bk$ in the Lippmann-Schwinger equation (\ref{lippmann}), see also the introduction. In fact Lemma \ref{lemlippschw} (iii) is to our knowledge the best one can say about the derivatives of the generalized eigenfunctions w.r.t. the coordinates. Note that the higher derivatives ($|\alpha|\geq 2$) become unbounded for small $k$. In \cite{panati:01} it is claimed that the derivatives stay bounded for small $k$, see Proposition 3.8. therein. The proof of this proposition apparently disregard the behavior of the coordinate derivatives of $k$.\erem

\section{The flux-across-surfaces theorem}\label{chflux}

The FAST (\ref{fluxi}) is a relation between a scattering state and its corresponding asymptote. As already emphasized it is important to establish the FAST with conditions only on the scattering state (and the potential $V$). Since $\psi=\Omega_+\pout$ we get by the well known expansion Lemma \ref{lemexpansion} (ii)-(iv): $\psi(\bx,t)=\mathcal{F}_+^{-1}e^{-i\frac{k^2}{2}t}\pouth(\bk)$ and we can express the flux in (\ref{fluxi}) by its asymptote $\pouth(\bk)$. Therefore we will proceed in the following way: We will first prove a FAST under conditions on $\pouth(\bk)$ and then we will translate these conditions to the corresponding scattering state. The connection between a scattering state and its corresponding asymptote is given by the expansion Lemma \ref{lemexpansion} (ii) and (iii), cf. (\ref{mapi}).\footnote{Because of Lemma \ref{lemlippschw} (ii) we can use the generalized Fourier transform without the $\limmean$ whenever $\psi\in L^1(\RRR^3)$.}\label{fotl1} That means, as already emphasized in the introduction, that the properties of $\pouth$ are via (\ref{mapi}) inherited by the properties of the generalized eigenfunctions, which are in general very poor, see Lemma \ref{lemlippschw}, especially (iii). More precisely, we will see later in the mapping Lemma \ref{lemmapping} that the decay properties (i.e. for large $k$) of $\pouth(\bk)$ and its derivatives depend mostly on the differentiability of $\psi(\bx),$ while the properties of $\pouth(\bk)$ and its derivatives for small $k$ are closely related to the corresponding properties of the generalized eigenfunctions $\varphi^*_+(\bx,\bk)$. Therefore we now define a class of asymptotes, $\ghut$, for which we can prove the FAST and which has the same poor properties for small $k$ as the generalized eigenfunctions in Lemma \ref{lemlippschw}. The exponents which determine the decay for large $k$ are optimized to get a large class and are of technical  interest. The class $\ghut$ of the suitable asymptotes is defined as follows: (In the following definition we have the Fourier transform of $\pout=\Omega_+^{-1}\psi$ (cf. (\ref{defout})) in mind.)

\bde\label{defgh} A function $f:\RRR^3\setminus\{0\}\to\CCC$ is in $\ghut$
if there is a constant $C\in\RRR_+$ with:
$$|f(\bk)|\leq C\lkr^{-15},$$ $$\left|\partial^{\alpha}_{\bk}f(\bk)\right|\leq C\lkr^{-6},\;|\alpha|=1,$$ $$\left|\kfac\:\partial^{\alpha}_{\bk}f(\bk)\right|\leq C\lkr^{-5},\;|\alpha|=2,\;\kfac=\frac{k}{\lkr}$$  $$\left|\frac{\partial^2}{\partial k^2}f(\bk)\right|\leq C\lkr^{-3}.$$ \ede
With that class we can formulate a FAST under conditions on $\pouth(\bk)$:
\bthe\label{theflux1} Let the potential satisfy the condition
$(V)_4$ and let zero be neither a resonance nor an eigenvalue of
$H$. Let $\pouth(\bk)\in\ghut$. Then
$\psi(\bx,t)=e^{-iHt}\Omega_+\psi_{\text{out}}(\bx)$ is
continuously differentiable except at the singularities of $V$ and
for any measurable $\Sigma\subset S^2$ and any $T\in\mathbb{R}$:
\begin{align}\label{flux1}
  \lim\limits_{R\to\infty}\int\limits_T^{\infty}\int\limits_{R\Sigma}\bj^{\psi}(\bx,t)
  \cdot d\bsig dt&=
  \lim\limits_{R\to\infty}\int\limits_T^{\infty}\int\limits_{R\Sigma}\left |\bj^{\psi}(\bx,t)
  \cdot d\bsig\right| dt=\int\limits_{C_{\Sigma}}|\pouth(\bk)|^2d^3k,
\end{align}
where $R\Sigma:=\{\bx\in\RRR^3:\bx=R\bom,\;\bom\in\Sigma\}$ and $C_{\Sigma}:=\{\bk\in\RRR^3:\be_k\in\Sigma\}$.
\ethe
The crucial condition in Theorem \ref{theflux1} is $\pouth(\bk)\in\ghut$. We introduce now the class $\mathcal{G}$ of scattering states for which we can prove that the corresponding asymptotes are in $\ghut$.

\bde\label{defg} $f$ is in $\gprime$ if \footnote{$C^8(H):=\bigcap\limits_{n=1}^{8}\dom(H^n)$}
$$f\in\mathcal{H}_{\text{a.c.}}(H)\cap C^8(H),$$
$$\lxr^2 H^nf\in\ltr,\;n\in\{0,1,2,...,8\},$$ $$\lxr^4 H^nf\in\ltr,\;n\in\{0,1,2,3\}.$$ Then $\mathcal{G}:=\bigcup\limits_{t\in\RRR}e^{-iHt}\gprime.$\ede
That means $\mathcal{G}$ is a subset of $\mathcal{H}_{\text{a.c.}}(H)$ and is invariant under finite time shifts, i.e. if $f\in\mathcal{G}$ then
$e^{-iHt}f\in\mathcal{G},\;\forall t\in\RRR.$  Furthermore $\mathcal{G}$ is dense in $\mathcal{H}_{\text{a.c.}}(H)$ which can be seen e.g. by the results used in \cite{amrein2:97}, p. 5368: Let $\mathcal{D}_{4}:=\{g(H)\langle x\rangle^{-4}\psi|g\in C_0^{\infty}(]0,\infty[),\psi\in\ltr\}.$ Since our potentials have no positive eigenvalues (cf. Section \ref{chsetup}) we have that $\mathcal{D}_{4}\subseteq\mathcal{H}_{\text{a.c.}}(H)$. It is easy to check that $\mathcal{D}_{4}$ is dense in $\mathcal{H}_{\text{a.c.}}(H)$. Moreover (cf. \cite{amrein2:97}) we have that $\mathcal{D}_{4}\subseteq\dom(H)\cap\dom(\langle x\rangle^{4})$. Again by \cite{amrein2:97} $H\mathcal{D}_{4}\subseteq\mathcal{D}_{4}$ which implies that $\mathcal{D}_{4}\subseteq\mathcal{G}.$ Hence, $\mathcal{G}$ is dense in $\mathcal{H}_{\text{a.c.}}(H)$. Note that the condition $\psi\in\mathcal{G}$ can be formulated also more explicitly (cf. Remark \ref{remg}). \z{We wish to remark that the condition $\psi\in C^8(H)$ seems to be natural: Wave functions in thermal equilibrium are typically in $C^{\infty}(H),$ see \cite{tumulka:05}.}

With Definition \ref{defg} we can state now the important mapping lemma,

\blem\label{lemmapping} Let $V\in(V)_4$ and let zero be neither a resonance nor an eigenvalue of $H$. Then:
$$\psi(\bx)\in\mathcal{G}\Rightarrow\widehat{\Omega_+\psi}(\bk)=\widehat{\psi}_{\text{out}}(\bk)\in\ghut.$$\elem
The proof is adapted from \cite{duerr:03} and can be found in the
appendix. The lemma holds also for $\Omega_+$ replaced by $\Omega_-$ and $\pout$ by $\psi_{\text{in}}.$\footnote{It would be interesting to have similar mapping properties for $\Omega_\pm^{-1}.$}

Theorem \ref{theflux1} and Lemma \ref{lemmapping} give the following corollary, the FAST under conditions on the scattering state.

\bcor\label{corflux} Let $V\in(V)_4$ and let zero be neither a resonance nor an eigenvalue of $H$. Let $\psi\in\mathcal{G}.$ Then for any measurable $\Sigma\subset S^2$
and any $T\in\mathbb{R}$:
\begin{align*}
  \lim\limits_{R\to\infty}\int\limits_T^{\infty}\int\limits_{R\Sigma}\bj^{\psi}(\bx,t)
  \cdot d\bsig dt&=
  \lim\limits_{R\to\infty}\int\limits_T^{\infty}\int\limits_{R\Sigma}\left |\bj^{\psi}(\bx,t)
  \cdot d\bsig\right| dt=\int\limits_{C_{\Sigma}}|\pouth(\bk)|^2d^3k.
\end{align*}
\ecor

\brem\label{remg} Instead of the condition $\psi\in\mathcal{G}$ one can give of course also the condition on $\psi$ and $V$ more explicitly. In the following we will give two examples for $\psi$ and $V$ such that $\psi\in\gprime$. The set of wave functions $\mathcal{G}$ for which the FAST holds is then---according to Definition \ref{defg}---given by the set $$\mathcal{G}=\bigcup\limits_{t\in\RRR}e^{-iHt}\gprime.$$
Let $H^{m,s}$ the weighted Sobolev space 
\begin{align*}
 H^{m,s}:=\left\{f\in\ltr|\;(1+x^2)^{\frac{s}{2}}(1-\Delta)^{\frac{m}{2}}f\in\ltr\right\}.\end{align*}
Then one can find for example the following conditions for which $\psi\in\gprime.$ 
\begin{itemize}
\item[(i)] $V\in(V)_2$, $V\in C^{14}(\RRR^3\setminus\mathcal{E})$, where $\mathcal{E}$ denotes the set of singularities of $V$ and\\ $\psi\in \mathcal{H}_{\text{a.c.}}(H)\cap C_0^{16}(\RRR^3\setminus\mathcal{E}).$
\item[(ii)] $V\in(V)_2$, $V\in H^{14,2}\cap H^{4,4}$ and $\psi\in\mathcal{H}_{\text{a.c.}}(H)\cap H^{16,2}\cap H^{6,4}.$
\end{itemize}
Clearly both  sets for $\psi$ are dense in $\mathcal{H}_{\text{a.c.}}(H).$
\erem 
\bpro{ of Theorem \ref{theflux1}} We will prove the
flux-across-surfaces theorem (\ref{flux1}) for some $T>0.$ This is
sufficient since: ($\widetilde{T}\leq 0,\;T>0)$
\begin{align}\label{fluxtime}
  \lim\limits_{R\to\infty}\int\limits_{\widetilde{T}}^{\infty}\int\limits_{R\Sigma}\bj^{\psi}(\bx,t)
  \cdot d\bsig dt&=\lim\limits_{R\to\infty}\int\limits_{T}^{\infty}\int\limits_{R\Sigma}\bj^{\widetilde{\psi}}(\bx,t)
  \cdot d\bsig dt,
\end{align}
with (in the second line we use Lemma \ref{lemexpansion} (ii)-(iv), again without the $\limmean$, because of Lemma \ref{lemlippschw} (ii) and $\pouth(\bk)\in\ghut\subset L^1(\RRR^3)$)
\begin{align}
  \label{fluxrest}
  \widetilde{\psi}(\bx,t)&=\psi(\bx,t+\widetilde{T}-T)=(2\pi)^{-\frac{3}{2}}\int e^{-i\frac{k^2t}{2}}e^{i\frac{k^2(T-\widetilde{T})}{2}}\pouth(\bk)\phip(\bx,\bk)d^3k\notag\\&=:(2\pi)^{-\frac{3}{2}}\int e^{-i\frac{k^2t}{2}}\couth(\bk)\phip(\bx,\bk)d^3k.
\end{align}
It is easy to check that $\couth(\bk)\in\ghut,$ if
$\pouth(\bk)\in\ghut,$ which means that
$\ghut$ is invariant under finite time
shifts. Hence, With (\ref{fluxtime}) and (\ref{fluxrest}) we
get:
\begin{align*}
  \lim\limits_{R\to\infty}\int\limits_{\widetilde{T}}^{\infty}\int\limits_{R\Sigma}\bj^{\psi}(\bx,t)
  \cdot d\bsig dt&=\hspace{-1,2pt}\lim\limits_{R\to\infty}\int\limits_{T}^{\infty}\int\limits_{R\Sigma}\bj^{\widetilde{\psi}}(\bx,t)
  \cdot d\bsig dt=\hspace{-1,2pt}\int\limits_{C_{\Sigma}}|\couth(\bk)|^2d^3k=\hspace{-1,2pt}\int\limits_{C_{\Sigma}}|\pouth(\bk)|^2d^3k.
\end{align*}
Of course, this argument is also valid for the integration over
$|\bj^{\psi}(\bx,t)\cdot d\bsig|.$

Let $T>0$ be fixed. Using Lemma \ref{lemexpansion} (ii)-(iv) and (\ref{lippmann})
we get:
\begin{align}
  \label{eq:first}
  \psi(\bx,t)&=(2\pi)^{-\frac{3}{2}}\int e^{-i\frac{k^2t}{2}}\pouth(\bk)\phip(\bx,\bk)d^3k\notag\\&=:(2\pi)^{-\frac{3}{2}}\int e^{-i\frac{k^2t}{2}}\pouth(\bk)e^{i\bk\cdot\bx}d^3k+(2\pi)^{-\frac{3}{2}}\int e^{-i\frac{\bk^2t}{2}}\pouth(\bk)\eta(\bx,\bk)d^3k\notag\\&=:\alpha(\bx,t)+\beta(\bx,t).
\end{align}
The flux generated by this wave function is:
\begin{equation}
  \label{eq:flux1}
  \bj^{\psi}(\bx,t)=\operatorname{Im}(\alpha^*\nabla\alpha+\alpha^*\nabla\beta+\beta^*\nabla\alpha+\beta^*\nabla\beta),
\end{equation}
where $\alpha$ is obviously continuously differentiable and for the
differentiability of $\beta$ see \cite{teufel2:99}, (20) and
(28)-(30). In \cite{daumer:96} and \cite{teufel2:99} the function $\alpha(\bx,t)$ is estimated using the formula
\begin{equation}
  \alpha(\bx,t)=(2\pi i t)\int e^{i\frac{|\bx-\by|^2}{2t}}\pout(\by)d^3y
\end{equation}
and conditions on $\pout(\bx).$ According to Lemma \ref{lemmapping} we can control $\pouth(\bk)$, but not $\pout(\bx).$ Hence, we have to estimate $\alpha(\bx,t)$ directly in terms of $\pouth(\bk).$ This will be done by using stationary phase methods. First, we will calculate $\bj_0^{\psi}=\operatorname{Im}(\alpha^*\nabla\alpha)$ by using Lemma \ref{lemstatphas}, which is formulated for a special class of wave functions $\widehat{\mathcal{K}}\supset\ghut$. This set has similar weak conditions as the set $\ghut$ due to the necessarily poor properties of $\pouth(\bk)$ (see the discussion before Definition \ref{defg}). Again we give here optimized decay properties, which are, however, not that strong as in the case of $\ghut$. 
\bde\label{defk} A function $f:\RRR^3\setminus\{0\}\to\CCC$ is in $\widehat{\mathcal{K}}$
if there is a constant $C\in\RRR_+$ with:
$$|f(\bk)|\leq C\lkr^{-4},$$ $$\left|\partial^{\alpha}_{\bk}f(\bk)\right|\leq C,\;|\alpha|=1,\;\;\left|\kfac\partial^{\alpha}_{\bk}f(\bk)\right|\leq C\lkr^{-1},\;|\alpha|=2,$$ $$\left|\frac{\partial}{\partial k}f(\bk)\right|\leq C\lkr^{-1},\;\;\left|
\frac{\partial^2}{\partial k^2}f(\bk)\right|\leq C\lkr^{-2}.$$
\ede
With that class of wave functions we can formulate

\blem\label{lemstatphas} Let $\chi(\bk)$ be in
$\widehat{\mathcal{K}}.$ Then there exists a constant
$L\in\mathbb{R_+}$ so that for all
$\bx\in\mathbb{R}^3$ and $t\in\mathbb{R},\;t\neq 0$:
\begin{equation}\label{hoermander}
\left|\int
e^{-i\frac{k^2}{2}t+i\bk\cdot\bx}\chi(\bk)d^3k-\left(\frac{2\pi}{it}\right)^{\frac{3}{2}}
e^{i\frac{x^2}{2t}}\chi(\bks)\right|<\frac{L}{t^{2}},
\end{equation}
where $\bks=\frac{\bx}{t}.$ \elem  The proof of the lemma can be found in the appendix.

Applying that lemma on $\alpha(\bx,t)$ in (\ref{eq:first}) we get,
with an appropriately chosen constant $L$:
\begin{equation}
  \label{eq:ann1}
  \left|\alpha(\bx,t)-\left(\frac{1}{it}\right)^{\frac{3}{2}}
  e^{i\frac{x^2}{2t}}\pouth\left(\frac{\bx}{t}\right)\right|<\frac{L}{t^{2}}
\end{equation}
and analogously:
\begin{equation}
  \label{eq:ann2}
  \left|\nabla\alpha(\bx,t)-i\left(\frac{1}{it}\right)^{\frac{3}{2}}
  e^{i\frac{x^2}{2t}}\left(\frac{\bx}{t}\right)\pouth\left(\frac{\bx}{t}\right)\right|
  <\frac{L}{t^{2}},
\end{equation}
which gives for the flux
$\bj_0^{\psi}=\operatorname{Im}(\alpha^*\nabla\alpha)$:
\begin{equation}
  \label{eq:fluxfree}
  \left|\bj_0^{\psi}(\bx,t)-\left(\frac{1}{t}\right)^{3}\left(\frac{\bx}{t}\right)\left|\pouth\left(\frac{\bx}{t}\right)\right|^2\right|<\frac{L}{t^{\frac{7}{2}}}.
\end{equation}
We begin with the first term $\bj_0^{\psi}$ in (\ref{eq:flux1})
for times $t>R^{\frac{5}{6}}$: (We choose $R$ big enough, so that
$R^{\frac{5}{6}}>T.$)
\begin{equation}\label{longtime}
  \int\limits_{R^{\frac{5}{6}}}^{\infty}\int\limits_{\Sigma}\bj_0^{\psi}(R\bn,t)
  \cdot \bn R^2d\Omega dt.
\end{equation}
Inserting the asymptotic expression (\ref{eq:fluxfree}) for the
flux $\bj_0^{\psi}$ we get instead of (\ref{longtime}):
\begin{equation}\label{longtimea}
  \int\limits_{R^{\frac{5}{6}}}^{\infty}\int\limits_{\Sigma}\left|\pouth\left(\frac{R\bn}{t}\right)\right|^2\frac{R^3}{t^4}d\Omega dt=\int\limits_0^{R^{\frac{1}{6}}}\int\limits_{\Sigma}\left|\pouth\left(\bk\right)\right|^2k^2d\Omega dk,
\end{equation}
where we substituted $\bk:=\frac{R\bn}{t}.$
(\ref{longtimea}) gives in the limit already the right result:
\begin{equation}
  \lim\limits_{R\to\infty}\int\limits_0^{R^{\frac{1}{6}}}\int\limits_{\Sigma}|\pouth(\bk)|^2 k^2 d\Omega dk=\int\limits_{C_{\Sigma}}|\pouth(\bk)|^2 d^3k.
\end{equation}
From (\ref{eq:fluxfree})-(\ref{longtimea}) it is clear that also the modulus of
$\bj_0^{\psi}$ gives the right result. Hence, by justifying the
use of the asymptotic expression for $\bj_0^{\psi}$, showing that
the integral (\ref{longtime}) is negligible for times smaller than
$R^{\frac{5}{6}}$ (and large $R$) and by proving the smallness of
the contributions of the three other terms in (\ref{eq:flux1}) we
get Theorem \ref{theflux1}.

Using (\ref{eq:fluxfree}) we can estimate the error between
(\ref{longtime}) and (\ref{longtimea}):
\begin{equation}
  L\int\limits_{R^{\frac{5}{6}}}^{\infty}\int\limits_{\Sigma}R^2t^{-\frac{7}{2}} d\Omega dt=\frac{8\pi L}{5}R^{-\frac{1}{12}},
\end{equation}
 which tends to zero for large $R.$

We evaluate now the flux integral for times smaller than
$R^{\frac{5}{6}}:$
\begin{equation}
  \int\limits_T^{R^{\frac{5}{6}}}\int\limits_{\Sigma}\bj_0^{\psi}(\bn R,t)
  \cdot \bn R^2d\Omega dt.
\end{equation}
Substituting $t\to R t$ we get:
\begin{equation}\label{psi0}
  \left|\int\limits_{\frac{T}{R}}^{R^{-\frac{1}{6}}}\int\limits_{\Sigma}\bj_0^{\psi}(R\bn ,tR)
  \cdot \bn R^3d\Omega dt\right|\leq\int\limits_{\frac{T}{R}}^{R^{-\frac{1}{6}}}\int\limits_{\Sigma}|\alpha(R\bn,tR)||\nabla_{\bx}\alpha(\bx,tR)|_{\bx=R\bn}R^3d\Omega dt.
\end{equation}
We estimate $\alpha$ and $\nabla\alpha$ separately. We start with
$\alpha$:
\begin{equation}\label{psi1}
  \alpha(R\bn,tR)=(2\pi)^{-\frac{3}{2}}\int e^{-it(\frac{k^2}{2}R-\bk\frac{R\bn}{t})}\pouth(\bk)d^3k
\end{equation}
The exponent of the $e$-function has the stationary point at
$\kstatb=\frac{1}{t}.$ Since $t\in[\frac{T}{R},R^{-\frac{1}{6}}]$,
$\kstatb\in[R^{\frac{1}{6}},\frac{R}{T}[.$ Big momenta should be
negligible, hence we divide the integration over $\bk$ in small
momenta up to $k<R^{\frac{1}{6}}$ and larger ones. This will be
done by the following functions:
\begin{equation}
  f_1(\bk)=\begin{cases}1\text{ for }k<\frac{1}{2}R^{\frac{1}{6}},\\\cos^2\left(\left(k-\frac{1}{2}R^{\frac{1}{6}}\right)\frac{\pi}{2}\right)\text{ for }\frac{1}{2}R^{\frac{1}{6}}\leq k\leq\frac{1}{2}R^{\frac{1}{6}}+1, \\0\text{ otherwise},\end{cases}
\end{equation}
\begin{equation}
  f_2(\bk)=\begin{cases}0\text{ for }k<\frac{1}{2}R^{\frac{1}{6}},\\\sin^2\left(\left(k-\frac{1}{2}R^{\frac{1}{6}}\right)\frac{\pi}{2}\right)\text{ for }\frac{1}{2}R^{\frac{1}{6}}\leq k\leq\frac{1}{2}R^{\frac{1}{6}}+1, \\1\text{ otherwise}.\end{cases}
\end{equation}
We have then $f_1(\bk)+f_2(\bk)\equiv 1$ and get for (\ref{psi1}):
\begin{align}\label{psi2}
  \alpha(R\bn,tR)=&(2\pi)^{-\frac{3}{2}}\int e^{-it(\frac{k^2}{2}R-\bk\frac{R\bn}{t})}\pouth(\bk)f_1(\bk)d^3k\notag\\&+(2\pi)^{-\frac{3}{2}}\int e^{-it(\frac{k^2}{2}R-\bk\frac{R\bn}{t})}\pouth(\bk)f_2(\bk)d^3k=:I_1+I_2.
\end{align}
We choose now $R$ large enough (such that
$\frac{1}{2}R^{\frac{1}{6}}>1$), which means that the first
integral in (\ref{psi2}) has no stationary point anymore. We will
do two integration by parts:
\begin{align}\label{psi3}
  I_1&=(2\pi)^{-\frac{3}{2}}\int e^{-it(\frac{k^2}{2}R-\bk\frac{R\bn}{t})}\pouth(\bk)f_1(\bk)d^3k\notag\\&=(2\pi)^{-\frac{3}{2}}\int \left(\nabla_{\bk}e^{-it(\frac{k^2}{2}R-\bk\frac{R\bn}{t})}\right)\cdot\frac{-i(Rt\bk-R\bn)}{|Rt\bk-R\bn|^2}\pouth(\bk)f_1(\bk)d^3k\notag\\&=-(2\pi)^{-\frac{3}{2}}\int e^{-it(\frac{k^2}{2}R-\bk\frac{R\bn}{t})}\left(\nabla_{\bk}\cdot\left(\frac{-i(Rt\bk-R\bn)}{|Rt\bk-R\bn|^2}\pouth(\bk)f_1(\bk)\right)\right)d^3k\notag\\&=:(2\pi)^{-\frac{3}{2}}\int e^{-it(\frac{k^2}{2}R-\bk\frac{R\bn}{t})}\left(\nabla_{\bk}\cdot\bg(\bk)\right)d^3k\notag\\&=(2\pi)^{-\frac{3}{2}}\int e^{-it(\frac{k^2}{2}R-\bk\frac{R\bn}{t})}\left(\nabla_{\bk}\cdot\left(\frac{-i(Rt\bk-R\bn)}{|Rt\bk-R\bn|^2}\left(\nabla_{\bk}\cdot\bg(\bk)\right)\right)\right)d^3k.
\end{align}
The gradient can be written as:
\begin{align}\label{psi4}
\nabla_{\bk}\cdot\left(\frac{-i(Rt\bk-R\bn)}{|Rt\bk-R\bn|^2}\left(\nabla_{\bk}\cdot\bg(\bk)\right)\right)&=\sum\limits_{i,j=1}^3\partial_{k_j}\left(\frac{-i(Rt\bk_j-R\bn_j)}{|Rt\bk-R\bn|^2}\left(\partial_{k_i}\bg_i(\bk)\right)\right).
\end{align}
A straightforward calculation yields for the right hand side of
(\ref{psi4}): (we consider one summand)
\begin{align}\label{psi5}
\left|\partial_{k_j}\left(\frac{-i(Rt\bk_j-R\bn_j)}{|Rt\bk-R\bn|^2}\left(\partial_{k_i}\bg_i(\bk)\right)\right)\right|\leq & C_1\frac{R^2t^2|\pouth(\bk)||f_1(\bk)|}{|Rt\bk-R\bn|^4}+C_2\frac{Rt|\partial_{k_i}(\pouth(\bk)f_1(\bk))|}{|Rt\bk-R\bn|^3}\notag\\&+C_3\frac{Rt|\partial_{k_j}(\pouth(\bk)f_1(\bk))|}{|Rt\bk-R\bn|^3}+C_4\frac{|\partial_{k_i}\partial_{k_j}(\pouth(\bk)f_1(\bk))|}{|Rt\bk-R\bn|^2},
\end{align}
with constants $C_k>0,\;k=1,2,3,4.$ 
Since $0\leq k<\frac{1}{2}R^{\frac{1}{6}}+1$ and $0<t\leq
R^{-\frac{1}{6}}$ we have:
\begin{align}\label{psi6}
|Rt\bk-R\bn|&\geq\frac{1}{2}R-R^{\frac{5}{6}}\geq\frac{1}{3}R,
\end{align}
if $R$ is large enough.
Using (\ref{psi6}) and the definition of $f_1(\bk)$ we find, with an appropriately chosen constant $M>0$, 
instead of (\ref{psi5}):
\begin{align}\label{psi7}
\left|\partial_{k_j}\left(\frac{-i(Rt\bk_j-R\bn_j)}{|Rt\bk-R\bn|^2}\left(\partial_{k_i}\bg_i(\bk)\right)\right)\right|\leq
&
\frac{Mt^2}{R^2}|\pouth(\bk)|\notag\\&+\frac{Mt}{R^2}\left(|\pouth(\bk)|+|\partial_{k_i}\pouth(\bk)|+|\partial_{k_j}\pouth(\bk)|\right)\notag\\&+\frac{M}{R^2}\left(|\partial_{k_j}\pouth(\bk)|+|\partial_{k_i}\pouth(\bk)|+|\partial_{k_i}\partial_{k_j}\pouth(\bk)|\right)\notag\\&+\frac{M}{R^2}\left(|\pouth(\bk)||\partial_{k_i}\partial_{k_j}f_1(\bk)|\right).
\end{align}
Using $|\partial_{\bk}^{\alpha}\pouth(\bk)|\leq
C\lkr^{-4},\;|\alpha|\leq 1,$ we get by (\ref{psi3}) and
(\ref{psi7}) an appropriate constant $M'>0$ with:
\begin{align}\label{psi8}
\left|I_1\right|\leq &
\frac{M'(t+1)^2}{R^2}+\frac{M(t+1)^2}{R^2}\int|\partial_{k_i}\partial_{k_j}\pouth(\bk)|k^2dkd\Omega\notag\\&+\frac{M C(t+1)^2}{R^2}\int\lkr^{-4}|\partial_{k_i}\partial_{k_j}f_1(\bk)|k^2dkd\Omega.
\end{align}
To integrate the second derivatives we use that
$|\kfac\partial_{\bk}^{\alpha}\pouth(\bk)|\leq
C\lkr^{-4},\;|\alpha|=2$ and\\
$k |\partial_{k_i}\partial_{k_j}f_1(\bk)|\leq C\lkr.$
Hence, with an appropriately chosen constant $C'$ we arrive at:
\begin{align}\label{psi9}
\left|I_1\right|\leq & \frac{C'(t+1)^2}{R^2}.
\end{align}
We estimate now $I_2.$ Since $\pouth(\bk)\in\ghut$
we have:
\begin{align}\label{psi11}
  \left|I_2\right|&\leq (2\pi)^{-\frac{3}{2}}C\int\limits_{k>\frac{1}{2}R^{\frac{1}{6}}}\lkr^{-15}d^3k\leq C'' R^{-2},\end{align}
with an appropriately chosen constant $C''>0.$
Hence, we find:
\begin{align}\label{psi12}
  \left|\alpha(R\bn,tR)\right|=&\left|I_1+I_2\right|\leq (C'+C'')(1+t)^2R^{-2}=:C'(1+t)^2R^{-2}.
\end{align}
In a similar way we can estimate $\nabla\alpha$ by:
\begin{align}\label{psi13}
  \left|\nabla_{\bx}\alpha(\bx,tR)\right|_{\bx=R\bn}&\leq C'(1+t)R^{-1}.
\end{align}
To get this estimate we split again the analogous integral to
(\ref{psi1}) into small and big momenta. The first part will be
estimated by one partial integration using that
$|\partial_{\bk}^{\alpha}\pouth(\bk)|\leq
C\lkr^{-5},\;|\alpha|\leq 1,$ and
$|\kfac\partial_{\bk}^{\alpha}\pouth(\bk)|\leq
C\lkr^{-5},\;|\alpha|=2,$ the second part (which is analogous to
(\ref{psi11})) by using that $|\pouth(\bk)|\leq C\lkr^{-10}.$
Inserting (\ref{psi12}) and (\ref{psi13}) into (\ref{psi0}) we
get:
\begin{align}\label{psi14}
  \int\limits_{\frac{T}{R}}^{R^{-\frac{1}{6}}}\int\limits_{\Sigma}|\alpha(R\bn,tR)||\nabla_{\bx}\alpha(\bx,tR)|_{\bx=R\bn}R^3d\Omega dt &\leq 4\pi C'^2\int\limits_0^{R^{-\frac{1}{6}}}(1+t)^3dt,
\end{align}
which tends to zero for $R\to\infty.$

It remains to show that the three other terms in (\ref{eq:flux1}) are negligible.
In \cite{teufel2:99} (Equations (15) and (16)) the function
$\beta(\bx,t)$ is estimated for some $R_0>0$ by:
\begin{equation}
  \label{eq:usualbeta1}
  \sup\limits_{\bx\in\Sigma_R}{|\beta(\bx,t)|\leq c\frac{1}{R(t+R)}},\;\forall R>0,
\end{equation}
\begin{equation}
  \label{eq:usualbeta2}
  \sup\limits_{\bx\in\Sigma_R}{|\nabla\beta(\bx,t)|\leq c\frac{1}{R(t+R)}},\;\forall R>R_0,
\end{equation}
for $t\geq T.$ The constant $c$ depends on $T$, $\pouth(\bk)$ and
$\frac{\partial}{\partial k}\pouth(\bk),$ and is finite for
$\pouth(\bk)\in\ghut$ (cf. (20)-(28) in \cite{teufel2:99}). It is also shown that the
last term in (\ref{eq:flux1}) is negligible (cf. p. 10 in \cite{teufel2:99}).
In \cite{teufel2:99} there are also estimates on the
$\alpha(\bx,t)$ terms, but not under the conditions which we must
require. We start with the second term in (\ref{eq:flux1}):
\begin{align}\label{psi15}
  \left|\int\limits_T^{\infty}\int\limits_{\Sigma}\operatorname{Im}(\alpha^*\nabla\beta) R^2 \bn d\Omega dt\right|&\leq\int\limits_T^{\infty}\int\limits_{\Sigma}\left|\alpha\right|\left|\nabla\beta\right| R^2 d\Omega dt\leq\int\limits_0^{\infty}\int\limits_{\Sigma}\left|\alpha\right|\frac{c}{R(t+R)}R^2 d\Omega dt.
\end{align}
We divide again the time integration into two parts:
\begin{align}\label{psi16}
 \int\limits_0^{\infty}\int\limits_{\Sigma}\left|\alpha\right|\frac{c}{R(t+R)}R^2 d\Omega dt&=\int\limits_0^{R^{\frac{5}{6}}}\int\limits_{\Sigma}\left|\alpha\right|\frac{c}{R(t+R)}R^2 d\Omega dt+\int\limits_{R^{\frac{5}{6}}}^{\infty}\int\limits_{\Sigma}\left|\alpha\right|\frac{c}{R(t+R)}R^2 d\Omega dt.
\end{align}
Hence, with (\ref{psi12}) the first part is:
\begin{align}\label{psi17}
 \int\limits_0^{R^{\frac{5}{6}}}\int\limits_{\Sigma}\left|\alpha(R\bn,t)\right|\frac{c}{R(t+R)}R^2 d\Omega
 dt&=\int\limits_0^{R^{-\frac{1}{6}}}\int\limits_{\Sigma}\left|\alpha(R\bn,tR)\right|\frac{c}{R^2(1+t)}R^3 d\Omega
 dt\notag\\&\leq\int\limits_0^{R^{-\frac{1}{6}}}\int\limits_{\Sigma}\frac{C'c(1+t)}{R} d\Omega dt,
\end{align}
which tends to zero for $R\to\infty.$

It remains the second term in (\ref{psi16}). Applying the asymptotic
expression (\ref{eq:ann1}) for $\alpha$ we get:
\begin{align}\label{psi19}
 \int\limits_{R^{\frac{5}{6}}}^{\infty}\int\limits_{\Sigma}\left|\alpha(R\bn,t)\right|\frac{cR^2}{R(t+R)}d\Omega dt\leq &\int\limits_{R^{\frac{5}{6}}}^{\infty}\int\limits_{\Sigma}\left(\frac{1}{t}\right)^{\frac{3}{2}}\left|\pouth\left(\frac{R\bn}{t}\right)\right|\frac{cR^2}{R(t+R)}d\Omega dt\notag\\&+\int\limits_{R^{\frac{5}{6}}}^{\infty}\int\limits_{\Sigma}\frac{L}{t^2}\frac{cR^2}{R(t+R)}d\Omega dt\notag\\\leq &\frac{4\pi c}{\sqrt{R}}\int\limits_0^{R^{\frac{1}{6}}}\left|\pouth\left(\bk\right)\right|\frac{1}{\sqrt{k}}dk+4\pi c L R^{-\frac{5}{6}},
\end{align}
where we substituted $\bk:=\frac{R\bn}{t}.$
Since $\pouth\in\ghut$ the bound in (\ref{psi19})
is  finite  and tends to zero for $R\to\infty.$
The third term in (\ref{eq:flux1}) can be treated analogously to (\ref{psi15})-(\ref{psi19}). \epro

\section{Appendix}

\bpro{ of Lemma \ref{lemlippschw}} Lemma  \ref{lemlippschw} is
proven---following the idea of Ikebe \cite{ikebe:60}---in
\cite{teufel2:99}. The latter however contains a mistake
concerning the assertion (iii), which overlooked the need for the
smoothing factor $\kfac=\frac{k}{1+k},$ which puts the higher
derivatives of the generalized eigenfunctions into the ``right''
Banach space. The need for this  smoothing factor arises from the
derivative of $k$ which appears in the spherical wave part in
(\ref{lippmann}), see also the remarks in the introduction.
Observing that, the proof goes through verbatim. Our statement
(iv) follows also from the proof in \cite{teufel2:99}, replacing
coordinate derivatives by the derivatives after $k$. In this case
we note that there is no need for any smoothing factor. 
\epro\\\bpro{ of Lemma \ref{lemmapping}} Let $\psi\in\mathcal{G}.$ Then
there is a $\chi\in\gprime$ and a $t\in\RRR$
with:
\begin{equation*}
\psi=e^{-iHt}\chi.
\end{equation*}
Using the intertwining property
(\ref{inter}) we get:
\begin{equation}\label{map0}
\pout=\Omega_+^{-1}\psi=\Omega_+^{-1}e^{-iHt}\chi=e^{-iH_0t}\Omega_+^{-1}\chi=e^{-iH_0t}\cout.
\end{equation}
Since $\ghut$ is invariant under multiplication by $e^{-i\frac{k^2}{2}t}$ it
suffices to show that $\couth(\bk)$ is in $\ghut.$
Let $\chi\in\gprime.$ Since $\lxr^2H^n\chi(\bx)\in
\ltr,\;0\leq n\leq 8$ and $\lxr^4H^n\chi(\bx)\in
\ltr,\;0\leq n\leq 3$ we have:
\begin{align}\label{l1l2}
\begin{split}
H^n\chi(\bx)&\in L_1(\RRR^3)\cap \ltr,\;0\leq n\leq 8,\\
\lxr^jH^n\chi(\bx)&\in L_1(\RRR^3)\cap \ltr,\;0\leq n\leq
3,\;j=\{1,2\}.
\end{split}
\end{align}
Using the intertwining property (\ref{inter}) and Lemma
\ref{lemexpansion} (ii), (iii) (cf. (\ref{mapi}) and Footnote \ref{fotl1}) we have:
\begin{align}\label{map}
\frac{k^2}{2}\couth(\bk)&=\widehat{H_0\chi}_{\text{out}}(\bk)=\mathcal{F}(H_0\Omega_+^{-1}\chi)(\bk)=\mathcal{F}(\Omega_+^{-1}H\chi)(\bk)\notag\\&=(2\pi)^{-\frac{3}{2}}\int\varphi^*_+(\bx,\bk)(H\chi)(\bx)d^3x.
\end{align}
Applying $H_0$ $n$ times  on $\couth(\bk)$ ($0\leq n\leq 8$) we get:
\begin{align}\label{map1}
\frac{k^{2n}}{2^n}\couth(\bk)&=(2\pi)^{-\frac{3}{2}}\int\varphi^*_+(\bx,\bk)(H^n\chi)(\bx)d^3x.
\end{align}
Since the generalized eigenfunctions are bounded (Lemma
\ref{lemlippschw} (ii)) and $H^n\chi\in L_1(\RRR^3),\;0\leq n\leq
8$ we have with an appropriate constant $C$:
\begin{equation}\label{map2}
|\couth(\bk)|\leq C\lkr^{-16}\leq C\lkr^{-15}.
\end{equation}
Because of Lemma \ref{lemlippschw} (iii) and (\ref{l1l2}) we can
differentiate $\couth(\bk)$ w.r.t. the coordinates and get an
appropriate constant $C$ with:
\begin{equation}\label{map5a}
\left|\partial_{k_i}\couth(\bk)\right|=\left|(2\pi)^{-\frac{3}{2}}\int\left(\partial_{k_i}\varphi^*_+(\bx,\bk)\right)\chi(\bx)d^3x\right|\leq C,\;\forall\bk\in\RRR^3\setminus\{0\}.
\end{equation}
Applying $H_0$ three times in (\ref{map}) and differentiating w.r.t. $k_i$ we get similarly to (\ref{map5a}):
\begin{equation}\label{map5}
k^6\partial_{k_i}\couth(\bk)=8(2\pi)^{-\frac{3}{2}}\int\left(\partial_{k_i}\varphi^*_+(\bx,\bk)\right)(H^3\chi)(\bx)d^3x-6k^5\couth(\bk)\frac{k_i}{k}.
\end{equation}
Again the right hand side is bounded because of Lemma
\ref{lemlippschw} (iii), (\ref{l1l2}) and (\ref{map2}). Hence, we
get together with (\ref{map5a}):
\begin{equation}\label{map6}
\left|\partial_{k_i}\couth(\bk)\right|\leq C\lkr^{-6},\;\forall\bk\in\RRR^3\setminus\{0\}.
\end{equation}
To control a second derivative with respect to the coordinates we
have to multiply by the factor $\kfac$, since then the derivatives
of the generalized eigenfunctions $\varphi_{\pm}$ are bounded by
$c\lxr^2,$ see Lemma \ref{lemlippschw} (iii). Hence by
(\ref{l1l2}):
\begin{equation}\label{map6a}
\left|\kfac\partial_{k_j}\partial_{k_i}\couth(\bk)\right|=\left|8(2\pi)^{-\frac{3}{2}}\int\left(\kfac\partial_{k_j}\partial_{k_i}\varphi^*_+(\bx,\bk)\right)\chi(\bx)d^3x\right|\leq C,\;\forall\bk\in\RRR^3\setminus\{0\}.
\end{equation}
Using (\ref{map5}) we get:
\begin{align}\label{map6b}
k^6\kfac\partial_{k_j}\partial_{k_i}\couth(\bk)=&8(2\pi)^{-\frac{3}{2}}\int\left(\kfac\partial_{k_j}\partial_{k_i}\varphi^*_+(\bx,\bk)\right)(H^3\chi)(\bx)d^3x-30k^4\frac{k_j}{k}\frac{k_i}{k}\kfac\couth(\bk)\notag\\&-6k^5\frac{k_i}{k}\kfac\partial_{k_j}\couth(\bk)-6k^5\couth(\bk)\kfac\frac{k\delta_{ij}k-k_ik_j}{k^3}-6k^5\frac{k_j}{k}\kfac\partial_{k_i}\couth(\bk),
\end{align}
 where the right hand side is bounded because of Lemma \ref{lemlippschw} (iii), (\ref{l1l2}), (\ref{map1}) and (\ref{map6}). Hence:
\begin{equation}\label{map7}
\left|\kfac\partial_{\bk}^{\alpha}\couth(\bk)\right|\leq
C\lkr^{-6}\leq C\lkr^{-5},\;|\alpha|=2,\;\forall\bk\in\RRR^3\setminus\{0\}.
\end{equation}
(\ref{map6}) implies also:
\begin{equation}\label{mapb}
\left|\partial_{k}\couth(\bk)\right|\leq
C\lkr^{-6},\;\forall\bk\in\RRR^3\setminus\{0\}.
\end{equation}
Applying $H_0$ two times in (\ref{map}) and differentiating two times w.r.t. $k$ we get by Lemma \ref{lemlippschw} (iv), (\ref{l1l2}), (\ref{map2}) and (\ref{mapb}) analogously to (\ref{map7}):
\begin{equation}\label{mapc}
\left|\partial_{k}^2\couth(\bk)\right|\leq
C\lkr^{-4}\leq
C\lkr^{-3},\;\forall\bk\in\RRR^3\setminus\{0\},
\end{equation}
which means that $\couth(\bk)\in\ghut.$ \epro

\bpro{ of Lemma \ref{lemstatphas}} At first sight Lemma \ref{lemstatphas} looks like a standard stationary phase result, e.g. Theorem 7.7.5 in \cite{hoermander:83}. But in our case we have (by necessity)
very weak conditions on the function $\chi(\bk),$ since we need to
use the lemma for $\chi(\bk)=\pouth(\bk)$.  Especially the second
derivative of $\chi(\bk)$ w.r.t. the coordinates becomes unbounded for $k\to 0.$ Furthermore, the stationary point $\bks$ is moving
with $\bx$ and $t$.

First, we extract the leading order term of the integral
(\ref{hoermander})
\begin{align}\label{extract}
\int e^{-i\frac{k^2}{2}t+i\bk\cdot\bx}\chi(\bk)d^3k=&\int
e^{-i\frac{k^2}{2}t+i\bk\cdot\bx}\left(\chi(\bk)-\chi(\bks)+\chi(\bks)\right)d^3k\notag\\=&\int
e^{-i\frac{k^2}{2}t+i\bk\cdot\bx}\chi(\bks)d^3k+\int
e^{-i\frac{k^2}{2}t+i\bk\cdot\bx}\left(\chi(\bk)-\chi(\bks)\right)d^3k.
\end{align}
The leading order term can be easily calculated:
\begin{eqnarray}\label{leading}
\int
e^{-i\frac{k^2}{2}t+i\bk\cdot\bx}\chi(\bks)d^3k=\left(\frac{2\pi}{it}\right)^{\frac{3}{2}}e^{i\frac{y^2}{2t}}\chi(\bks).
\end{eqnarray}
We will now calculate the error between the left hand side of
(\ref{extract}) and the leading order term (\ref{leading}):
\begin{equation}\label{extracta}
\int e^{-i\frac{k^2}{2}t+i\bk\cdot\bx}\left(\chi(\bk)-\chi(\bks)\right)d^3k.
\end{equation}
The following splitting of the integration area turns out to be convenient (cf. Figure 1):
\begin{align}\label{areas}
A_1:=&\{\bk\in\RRR^3:\kt=|\bk-\bks|<\frac{\ks}{2}\},\;A_2:=\{\bk\in\RRR^3:k<2\ks\wedge|\bk-\bks|\geq\frac{\ks}{2}\},\notag\\A_3:=&\{\bk\in\RRR^3:k\geq \frac{3}{2}\ks\}.
\end{align}
\begin{wrapfigure}{r}{5,3cm}
\centering
\includegraphics[width=4cm]{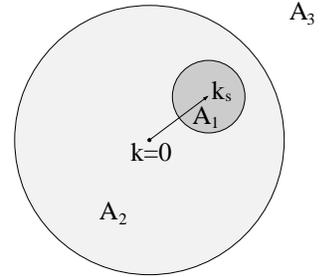}
\caption{Sketch of the three integration areas in the $\bk$-frame.}
\noindent \hrulefill
\label{test}
\end{wrapfigure}
The areas $A_2$ and $A_3$ have a small overlap. This is due to the use of suitable mollifiers. In $A_1$ and $A_2$ we shall perform two partial integrations w.r.t. the coordinates, in $A_3$ we shall perform the derivatives w.r.t. $k.$ 
Our proof will assume $\bx\neq 0.$ The case $\bx=0$ must be handled separately, but is much easier than the proof we give. It can be done by two partial integrations w.r.t. $k$ similarly to our procedure which handles the area $A_3$ (\ref{areas}).

We first divide the integration area into $A_1\cup A_2$ and $A_3$ by using the mollifier $\rho(\bk):$
\begin{equation}\label{rho}
\rho(\bk)=
\begin{cases}
1,\text{ for }k<\frac{3}{2}\ks,\\e\exp\left(-\frac{1}{1-\frac{\left(k-\frac{3}{2}\ks\right)^2}{\left(\frac{\ks}{2}\right)^2}}\right),\text{ for }\frac{3}{2}\ks\leq k<2\ks,\\
0,\text{ for }k\geq 2\ks.
\end{cases}
\end{equation}
The mollifier has the following properties:
\begin{align}\label{molli}
&\supp \rho=A_1\cup A_2,\;|\rho(\bk)|\leq 1,\;|1-\rho(\bk)|\leq 1,\notag\\
&\text{There is an }M>0\text{ such that: }|\partial_k\rho(\bk)|,|\partial_{\bk}^{\alpha}\rho(\bk)|\leq\frac{M}{\ks},\;|\alpha|=1,\notag\\&\text{and }|\partial_k^2\rho(\bk)|,|\partial_{\bk}^{\alpha}\rho(\bk)|\leq\frac{M}{\ks^2},\;|\alpha|=2.
\end{align}
Using $\rho$ we can write for (\ref{extracta}):
\begin{align}\label{extractb}
\int e^{-i\frac{k^2}{2}t+i\bk\cdot\bx}\left(\chi(\bk)-\chi(\bks)\right)d^3k=&\int
e^{-i\frac{k^2}{2}t+i\bk\cdot\bx}\rho(\bk)\left(\chi(\bk)-\chi(\bks)\right)d^3k\notag\\&+\int
e^{-i\frac{k^2}{2}t+i\bk\cdot\bx}\left(1-\rho(\bk)\right)\left(\chi(\bk)-\chi(\bks)\right)d^3k\notag\\=:&I_{12}+I_3.
\end{align}
We start with the estimation of $I_{12}.$ We define: 
\begin{align}\label{deff}
f(\bk)&:=\rho(\bk)\left(\chi(\bk)-\chi(\bks)\right),\;\bkt:=\bk-\bks
\end{align}
and get with two partial integration w.r.t. to $\bk:$
\begin{align}\label{extractc}
\left|I_{12}\right|=&\left|\int e^{-it\left(\frac{k^2}{2}-\bk\cdot\bks\right)}f(\bk)d^3k\right|\notag\\=&\frac{1}{t}\left|\int \left(\nabla_{\bk}e^{-i\frac{k^2}{2}t+i\bk\cdot\bx}\right)\cdot\frac{\bkt}{\kt^2}f(\bk)d^3k\right|\notag\\=&\frac{1}{t}\left|\int e^{-i\frac{k^2}{2}t+i\bk\cdot\bx}\left(\frac{\bkt\cdot\nabla_{\bk}f(\bk)-f(\bk)}{\kt^2}\right)d^3k\right|\notag\\=&\frac{1}{t^2}\left|\int \left(\nabla_{\bk}e^{-i\frac{k^2}{2}t+i\bk\cdot\bx}\right)\cdot\frac{\bkt}{\kt^2}\left(\frac{\bkt\cdot\nabla_{\bk}f(\bk)-f(\bk)}{\kt^2}\right)d^3k\right|\notag\\=&\frac{1}{t^2}\left|\int e^{-i\frac{k^2}{2}t+i\bk\cdot\bx}\left(\frac{f(\bk)-\bkt\cdot\nabla_{\bk}f(\bk)}{\kt^4}+\frac{1}{\kt^4}\sum\limits_{|\alpha_1|+|\alpha_2|=2}\bkt^{\alpha_1}\bkt^{\alpha_2}\partial_{\bk}^{\alpha_1}\partial_{\bk}^{\alpha_2}f(\bk)\right)d^3k\right|\notag\\\leq &\frac{1}{t^2}\int \left|\frac{f(\bk)-\bkt\cdot\nabla_{\bk}f(\bk)}{\kt^4}\right|d^3k+\frac{1}{t^2}\int\left|\frac{1}{\kt^4}\sum\limits_{|\alpha_1|+|\alpha_2|=2}\bkt^{\alpha_1}\bkt^{\alpha_2}\partial_{\bk}^{\alpha_1}\partial_{\bk}^{\alpha_2}f(\bk)\right|d^3k.
\end{align}
Because of the definition of $\rho,$ the integration area in (\ref{extractc}) is $A_1\cup A_2$ (cf. (\ref{areas}) and (\ref{rho})). We will divide this area into $A_1$ and $A_2.$ Hence, $I_{12}$ is estimated by:
\begin{align}\label{extractd}
\left|I_{12}\right|\leq&\frac{1}{t^2}\int\limits_{A_1}\left|\frac{f(\bk)-\bkt\cdot\nabla_{\bk}f(\bk)}{\kt^4}\right|d^3k+\frac{1}{t^2}\int\limits_{A_1}\left|\frac{1}{\kt^4}\sum\limits_{|\alpha_1|+|\alpha_2|=2}\bkt^{\alpha_1}\bkt^{\alpha_2}\partial_{\bk}^{\alpha_1}\partial_{\bk}^{\alpha_2}f(\bk)\right|d^3k\notag\\&+\frac{1}{t^2}\int\limits_{A_2}\left|\frac{f(\bk)-\bkt\cdot\nabla_{\bk}f(\bk)}{\kt^4}\right|d^3k+\frac{1}{t^2}\int\limits_{A_2}\left|\frac{1}{\kt^4}\sum\limits_{|\alpha_1|+|\alpha_2|=2}\bkt^{\alpha_1}\bkt^{\alpha_2}\partial_{\bk}^{\alpha_1}\partial_{\bk}^{\alpha_2}f(\bk)\right|d^3k\notag\\=:&I_1+\frac{1}{t^2}\int\limits_{A_2}\left|\frac{f(\bk)-\bkt\cdot\nabla_{\bk}f(\bk)}{\kt^4}\right|d^3k+\frac{1}{t^2}\int\limits_{A_2}\left|\frac{1}{\kt^4}\sum\limits_{|\alpha_1|+|\alpha_2|=2}\bkt^{\alpha_1}\bkt^{\alpha_2}\partial_{\bk}^{\alpha_1}\partial_{\bk}^{\alpha_2}f(\bk)\right|d^3k\notag\\=:&I_1+I_2.
\end{align}
We estimate $I_1$ first. With (\ref{areas}) and (\ref{rho}) we see that for $\bk\in A_1$, $\rho(\bk)\equiv 1$ and thus we have for (\ref{deff}): $f(\bk)=\chi(\bk)-\chi(\bks).$ Using Taylors formula and then substituting $\bk$ by $\bkt$ (cf. (\ref{deff})) we get for the first term $I_1^1$ of $I_1:$
\begin{align}\label{extracte}
I_{1}^1=&\frac{1}{t^2}\int\limits_{A_1}\left|\frac{f(\bk)-\bkt\cdot\nabla_{\bk}f(\bk)}{\kt^4}\right|d^3k\notag\\=&\frac{1}{t^2}\int\limits_{A_1}\left|\frac{\chi(\bk)-\chi(\bks)-(\bk-\bks)\cdot\nabla_{\bk}\chi(\bk)}{(\bk-\bks)^4}\right|d^3k\notag\\=&\frac{1}{t^2}\int\limits_{A_1}\left|\frac{\sum\limits_{|\alpha_1|+|\alpha_2|=2}(\bk-\bks)^{\alpha_1}(\bk-\bks)^{\alpha_2}\partial_{\bk}^{\alpha_1}\partial_{\bk}^{\alpha_2}\chi(\bxi)}{2(\bk-\bks)^4}\right|d^3k\notag\\=&\frac{1}{t^2}\int\limits_{A_1}\left|\frac{\sum\limits_{|\alpha_1|+|\alpha_2|=2}\bkt^{\alpha_1}\bkt^{\alpha_1}\partial_{\bk}^{\alpha_1}\partial_{\bk}^{\alpha_2}\chi(\bxi)}{2\kt^4}\right|d^3\kt,
\end{align}
where $\bxi$ is a vector between $\bks$ and $\bk$. Hence we have $\xi>\frac{\ks}{2}.$ Using Definition \ref{defk}, i.e. that $\partial_{k_i}\partial_{k_j}\chi(\bk)\leq Ck^{-1},$ we get for (\ref{extracte}):
\begin{align}\label{extractf}
I_{1}^1\leq&\frac{9C}{2t^2}\int\limits_{A_1}\frac{1}{\kt^2\xi}d^3\kt<\frac{36\pi C}{\ks t^2}\int\limits_{A_1}d\kt=\frac{18\pi C}{t^2}.
\end{align}
The second term of $I_1$ can be estimated analogously: Instead of $\bxi$ we have $\bk=\bkt+\bks$ with $k>\frac{\ks}{2}.$ It follows that $I_1$ is of order $t^{-2}$ uniform in $\bks.$
The estimation of $I_2$ is very similar, but $\rho(\bk)\neq 1$ on $A_2$. We use the volume factor $d^3k$ integrated over $A_2.$ Hence, it suffices to show that the integrands of the two terms of $I_2$ are bounded by $\frac{L}{\ks^3}$ or by $\frac{L}{\ks^2k}$ with some constant $L>0$ uniform in $\bks.$
The first integrand is:
\begin{align}\label{i2a}
\left|\frac{f(\bk)-\bkt\cdot\nabla_{\bk}f(\bk)}{\kt^4}\right|\leq&\left|\frac{\rho(\bk)(\chi(\bk)-\chi(\bks))}{\kt^4}\right|+\left|\frac{\left|\nabla_{\bk}f(\bk)\right|}{\kt^3}\right|\notag\\\leq&\left|\frac{\chi(\bk)-\chi(\bks)}{\kt^4}\right|+\sum\limits_{i=1}^{3}\left|\frac{\left|\partial_{k_i}\chi(\bk)\right|}{\kt^3}\right|+\sum\limits_{i=1}^{3}\left|\frac{\left|(\chi(\bk)-\chi(\bks))\partial_{k_i}\rho(\bk)\right|}{\kt^3}\right|.
\end{align}
By mean value theorem there exists a $\bxi\in(\bks,\bk)$ with:
\begin{align}\label{i2b}
|\chi(\bk)-\chi(\bks)|=&\left|\nabla_{\bk}\chi(\bxi)\right||\bk-\bks|\leq Ck+C\ks,
\end{align}
since $\chi(\bk)\in\widehat{\mathcal{K}}$.
Using (\ref{i2b}), (\ref{molli}), $\bk\in A_2$ (which means: $k<2\ks$, $\kt\geq\frac{\ks}{2}$) as well as $\left|\partial_{k_i}\chi(\bk)\right|\leq C,\;i=\{1,2,3\}$ we obtain:
\begin{align}\label{i2c}
\left|\frac{f(\bk)-\bkt\cdot\nabla_{\bk}f(\bk)}{\kt^4}\right|\leq&\frac{32C}{\ks^3}+\frac{16C}{\ks^3}+\frac{24C}{\ks^3}+\frac{48CM}{\ks^3}+\frac{24CM}{\ks^3}.
\end{align}
Similarly we estimate the integrand of the second term of $I_2$ (\ref{extractd}). We pick one summand: ($|\alpha_1|+|\alpha_2|=2$)
\begin{align}\label{i2d}
\left|\frac{1}{\kt^4}\bkt^{\alpha_1}\bkt^{\alpha_1}\partial_{\bk}^{\alpha_1}\partial_{\bk}^{\alpha_2}f(\bk)\right|\leq&\left|\frac{1}{\kt^2}\partial_{\bk}^{\alpha_1}\partial_{\bk}^{\alpha_2}f(\bk)\right|\notag\\\leq&\frac{4\left|\chi(\bk)-\chi(\bks)\right|\left|\partial_{\bk}^{\alpha_1}\partial_{\bk}^{\alpha_2}\rho(\bk)\right|}{\ks^2}+\frac{4\left|\partial_{\bk}^{\alpha_1}\rho(\bk)\right|\left|\partial_{\bk}^{\alpha_2}\chi(\bk)\right|}{\ks^2}\notag\\&+\frac{4\left|\partial_{\bk}^{\alpha_2}\rho(\bk)\right|\left|\partial_{\bk}^{\alpha_1}\chi(\bk)\right|}{\ks^2}+\frac{4\left|\partial_{\bk}^{\alpha_1}\partial_{\bk}^{\alpha_2}\chi(\bk)\right|}{\ks^2}\notag\\\leq&\frac{8CM}{\ks^3}+\frac{4CM}{\ks^3}+\frac{8CM}{\ks^3}+\frac{4C}{\ks^2k}.
\end{align}

It remains to estimate $I_3$ (\ref{extractb}). We introduce a convergence factor $\rho_{\epsilon}(\bk):$
\begin{equation}\label{rhoeps}
\rho_{\epsilon}(\bk)=
\begin{cases}
1,\text{ for }k<\frac{1}{\epsilon},\\e^{-\left(k-\frac{1}{\epsilon}\right)^2},\text{ for }k\geq\frac{1}{\epsilon},
\end{cases}
\end{equation}
with $0<\epsilon<\min(\frac{1}{2\ks};1).$
Then we get for $I_3$ (\ref{extractb}):
\begin{align}\label{i3a}
I_3=&\int
e^{-i\frac{k^2}{2}t+i\bk\cdot\bx}\left(1-\rho(\bk)\right)\left(\chi(\bk)-\chi(\bks)\right)d^3k\notag\\=&\int
e^{-i\frac{k^2}{2}t+i\bk\cdot\bx}\left(1-\rho(\bk)\right)\left(1-\rho_{\epsilon}(\bk)+\rho_{\epsilon}(\bk)\right)\left(\chi(\bk)-\chi(\bks)\right)d^3k\notag\\=&\lim\limits_{\epsilon\to 0}\int
e^{-i\frac{k^2}{2}t+i\bk\cdot\bx}\left(1-\rho(\bk)\right)\left(1-\rho_{\epsilon}(\bk)+\rho_{\epsilon}(\bk)\right)\left(\chi(\bk)-\chi(\bks)\right)d^3k\notag\\=&\lim\limits_{\epsilon\to 0}\int
e^{-i\frac{k^2}{2}t+i\bk\cdot\bx}\left(1-\rho(\bk)\right)\rho_{\epsilon}(\bk)\left(\chi(\bk)-\chi(\bks)\right)d^3k\notag\\&+\lim\limits_{\epsilon\to 0}\int
e^{-i\frac{k^2}{2}t+i\bk\cdot\bx}\left(1-\rho(\bk)\right)\left(1-\rho_{\epsilon}(\bk)\right)\left(\chi(\bk)-\chi(\bks)\right)d^3k\notag\\=&\lim\limits_{\epsilon\to 0}\int
e^{-i\frac{k^2}{2}t+i\bk\cdot\bx}\left(1-\rho(\bk)\right)\rho_{\epsilon}(\bk)\left(\chi(\bk)-\chi(\bks)\right)d^3k\notag\\&+\lim\limits_{\epsilon\to 0}\int\limits_{k=\frac{1}{\epsilon}}^{\infty}
e^{-i\frac{k^2}{2}t+i\bk\cdot\bx}\left(1-\rho_{\epsilon}(\bk)\right)\left(\chi(\bk)-\chi(\bks)\right)d^3k,
\end{align}
since $1-\rho\equiv 1$ on $\supp(1-\rho_{\epsilon})$ (cf. (\ref{rho}) and (\ref{rhoeps})).
The last term in the last line of (\ref{i3a}) is zero (since $\chi(\bk)\in\widehat{\mathcal{K}}$ and by a standard Riemann-Lebesgue argument) and we get for $I_3$:
\begin{align}\label{i3b}
I_3=&\lim\limits_{\epsilon\to 0}\int
e^{-it\left(\frac{k^2}{2}-\bk\cdot\bks\right)}\left(1-\rho(\bk)\right)\rho_{\epsilon}(\bk)\left(\chi(\bk)-\chi(\bks)\right)d^3k\notag\\=:&\lim\limits_{\epsilon\to 0}\int
e^{-it\left(\frac{k^2}{2}-\bk\cdot\bks\right)}f_{\epsilon}(\bk,\bks)k^2dkd\Omega.
\end{align}
We will perform now two partial integrations w.r.t. $k$:
\begin{align}\label{i3c}
|I_3|= &\left|\lim\limits_{\epsilon\to 0}\frac{1}{t}\int
e^{-it\left(\frac{k^2}{2}-\bk\cdot\bks\right)}\partial_k\left(\frac{k^2f_{\epsilon}(\bk,\bks)}{k-\be_k\cdot\bks}\right)dkd\Omega \right|\notag\\=&\left|\lim\limits_{\epsilon\to 0}\frac{1}{t^2}\int
e^{-it\left(\frac{k^2}{2}-\bk\cdot\bks\right)}\partial_k\left(\frac{1}{k-\be_k\cdot\bks}\partial_k\left(\frac{k^2f_{\epsilon}(\bk,\bks)}{k-\be_k\cdot\bks}\right)\right)dkd\Omega\right|\notag\\\leq&\frac{1}{t^2}\lim\limits_{\epsilon\to 0}\int\left|\partial_k\left(\frac{1}{k-\be_k\cdot\bks}\partial_k\left(\frac{k^2f_{\epsilon}(\bk,\bks)}{k-\be_k\cdot\bks}\right)\right)\right|dkd\Omega\notag\\=:&\frac{1}{t^2}\lim\limits_{\epsilon\to 0}\int\limits_{k\geq\frac{3}{2}\ks} |D|dkd\Omega\notag\\\leq&\frac{1}{t^2}\lim\limits_{\epsilon\to 0}\int\limits_{\frac{3}{2}\ks\leq k<2\ks} |D|dkd\Omega+\frac{1}{t^2}\lim\limits_{\epsilon\to 0}\int\limits_{2\ks\leq k<\frac{1}{\epsilon}} |D|dkd\Omega+\frac{1}{t^2}\lim\limits_{\epsilon\to 0}\int\limits_{k\geq\frac{1}{\epsilon}} |D|dkd\Omega\notag\\=:&I_3^1+I_3^2+I_3^3.
\end{align}
We start with the estimation of $I_3^1.$ Because of the integration area it suffices to show that $D$ is of order $\ks^{-1}.$ Since $\rho_{\epsilon}(\bk)\equiv 1$ in this area, $D$ is given by: ($(\cdot)'$ denotes the derivative w.r.t. $k$)
\begin{align}\label{i3d}
|D|\leq&\left|\frac{k^2}{(\kks)^2}\right|\left|\left((1-\rho(\bk))(\chi(\bk)-\chi(\bks))\right)''\right|\notag\\&+\left|\left(\frac{k^2}{(\kks)^2}\right)'+\frac{1}{\kks}\left(\frac{k^2}{\kks}\right)'\right|\left|\left((1-\rho(\bk))(\chi(\bk)-\chi(\bks))\right)'\right|\notag\\&+\left|\left(\frac{k^2}{\kks}\right)''\frac{1}{\kks}+\left(\frac{1}{\kks}\right)'\left(\frac{k^2}{\kks}\right)'\right|\cdot\notag\\&\hspace{6,5cm}\cdot\left|(1-\rho(\bk))(\chi(\bk)-\chi(\bks))\right|.
\end{align}
We shall use (sometimes in a slightly modified version):
\begin{align}\label{substi}
\frac{k^2}{(\kks)^2}&\leq\frac{k^2}{(k-\ks)^2}=\frac{(k-\ks+\ks)^2}{(k-\ks)^2}\leq 9, \text{ for } k\geq\frac{3}{2}\ks.
\end{align}
Using (\ref{substi}) we get instead of (\ref{i3d}):
\begin{align}\label{i3e}
|D|\leq & 9\left|\left((1-\rho(\bk))(\chi(\bk)-\chi(\bks))\right)''\right|+\frac{39}{k-\ks}\left|\left((1-\rho(\bk))(\chi(\bk)-\chi(\bks))\right)'\right|\notag\\&+\frac{47}{(k-\ks)^2}\left|(1-\rho(\bk))(\chi(\bk)-\chi(\bks))\right|.
\end{align}
Using $\chi(\bk)\in\widehat{\mathcal{K}}$, i.e.
\begin{align}\label{i3f}
\left|(\chi(\bk)-\chi(\bks))'\right|\leq & C\lkr^{-1}\leq C,\notag\\\left|(\chi(\bk)-\chi(\bks))''\right|\leq C\lkr^{-2}\leq & C\lkr^{-1}\leq C(1+\ks)^{-1},
\end{align}
since $k>\ks$, (\ref{molli}), (\ref{i2b}) and (\ref{substi}) we find that:
\begin{align}\label{i3f1}
|D|\leq & \frac{818CM}{\ks}+\frac{9C}{1+\ks}.
\end{align}
It follows that $I_3^1$ is of order $t^{-2}$ uniform in $\bks.$
It remains to estimate $I_3^2$ and $I_3^3.$ First we consider ``large'' $\ks:$ Let $2\ks\geq 1.$ $D$ on the integration area of $I_3^2$ (where $\frac{1}{\epsilon}>k\geq 2\ks$) is bounded by: (we use again $\left|\chi'(\bk)\right|\leq C\lkr^{-1}$ and $\left|\chi''(\bk)\right|\leq C\lkr^{-2}$)
\begin{align}\label{i3g}
|D|\leq&\left|\frac{k^2}{(\kks)^2}\right|\left|\chi''(\bk)\right|\notag\\&+\left|\left(\frac{k^2}{(\kks)^2}\right)'+\frac{1}{\kks}\left(\frac{k^2}{\kks}\right)'\right|\left|\chi'(\bk)\right|\notag\\&+\left|\left(\frac{k^2}{\kks}\right)''\frac{1}{\kks}+\left(\frac{1}{\kks}\right)'\left(\frac{k^2}{\kks}\right)'\right|\left|\chi(\bk)-\chi(\bks)\right|\notag\\\leq &\frac{4C}{\lkr^{2}}+\frac{10C}{(k-\ks)\lkr}+\frac{52C}{(k-\ks)^2},
\end{align}
where we used (\ref{substi}) (and analogous estimates) with $k\geq 2\ks.$
Inserting (\ref{i3g}) into (\ref{i3b}) we get:
\begin{align}\label{i3h}
|I_3^2|=&\frac{1}{t^2}\lim\limits_{\epsilon\to 0}\int\limits_{2\ks\leq k<\frac{1}{\epsilon}}|D|dkd\Omega\leq\frac{1}{t^2}\lim\limits_{\epsilon\to 0}\int\limits_{2\ks\leq k}|D|dkd\Omega=\frac{1}{t^2}\int\limits_{2\ks\leq k}|D|dkd\Omega,
\end{align}
which is integrable uniformly in $\bks$ for $2\ks\geq 1.$ Hence, $I_3^2$ is of order $t^{-2}$ uniformly in $\bks$ for $2\ks\geq 1.$ Similar we can estimate $I_3^3$ since then we have for $D$:
\begin{align}\label{i3i}
|D|\leq&4\left|\left(\rho_{\epsilon}(\bk)(\chi(\bk)-\chi(\bks))\right)''\right|+\frac{10}{k-\ks}\left|\left(\rho_{\epsilon}(\bk)(\chi(\bk)-\chi(\bks))\right)'\right|\notag\\&+\frac{26}{(k-\ks)^2}\left|\rho_{\epsilon}(\bk)(\chi(\bk)-\chi(\bks))\right|\notag\\\leq&4\left|\left(\rho_{\epsilon}(\bk)(\chi(\bk)-\chi(\bks))\right)''\right|+20\left|\left(\rho_{\epsilon}(\bk)(\chi(\bk)-\chi(\bks))\right)'\right|\notag\\&+104\left|\rho_{\epsilon}(\bk)(\chi(\bk)-\chi(\bks))\right|.
\end{align}
The integration of (\ref{i3i}) over the area $k\geq\frac{1}{\epsilon}$ yields a uniform bound in $\bks$ and $\epsilon$ for $2\ks\geq 1.$ 
It remains to estimate $I_3^2$ and $I_3^3$ for $2\ks<1.$ 
$I_3^3$ can be estimated analogous to (\ref{i3i}) since $k\geq\frac{1}{\epsilon}>1$ and we have again: 
\begin{align}\label{i3j}
\frac{1}{k-\ks}<\frac{1}{1-\ks}<2.
\end{align}
For $I_3^2$ we split the integration into:
\begin{align}\label{i3k}
I_3^2\leq&\frac{1}{t^2}\lim\limits_{\epsilon\to 0}\int\limits_{2\ks\leq k<1} |D|dkd\Omega+\frac{1}{t^2}\lim\limits_{\epsilon\to 0}\int\limits_{1\leq k<\frac{1}{\epsilon}} |D|dkd\Omega.
\end{align}
The second term of the right side of (\ref{i3k}) can be estimated analogous to $I_3^2$ for $2\ks\geq 1.$ 
Thus remains the following integral:
\begin{align}\label{i3l}
\frac{1}{t^2}\lim\limits_{\epsilon\to 0}\int\limits_{2\ks\leq k<1} |D|dkd\Omega.
\end{align}
The integrand is bounded by: (we use again (\ref{substi}))
\begin{align}\label{i3m}
|D|\leq&\left|\frac{k^2}{(\kks)^2}\right|\left|\chi''(\bk)\right|+\left|\left(\frac{k^2}{(\kks)^2}\right)'\right|\left|\chi'(\bk)\right|\notag\\&+\left|\left(\left(\frac{k^2}{\kks}\right)'\frac{1}{\kks}(\chi(\bk)-\chi(\bks))\right)'\right|\notag\\\leq&4C+\left|\left(\frac{k^2}{(\kks)^2}\right)'\right|C+\left|\left(\left(\frac{k^2}{\kks}\right)'\frac{1}{\kks}(\chi(\bk)-\chi(\bks))\right)'\right|\notag\\=:&|D_1|+|D_2|+|D_3|.
\end{align}
We have to integrate $D$ over a bounded interval. Hence, $D_1$ yields a uniform constant. The derivative in $D_2$ has at most two zeros in $A_3.$ So we can divide the integration area  into three subsets on which $\partial_k\left(\frac{k^2}{(\kks)^2}\right)$ does not change the sign. Then we can apply the fundamental theorem of calculus to conclude that also the second term yields a uniform constant, using (\ref{substi}).
$D_3$ can be written as:
\begin{align}\label{i3n}
|D_3|=&\left|\left(\left(\frac{k^2}{\kks}\right)'\frac{1}{\kks}(\chi(\bk)-\chi(0)-k\chi'(0)+\chi(0)-\chi(\bks)+k\chi'(0))\right)'\right|\notag\\=&\left|\left(\left(\frac{k^2}{\kks}\right)'\frac{1}{\kks}(\chi(\bk)-\chi(0)-k\chi'(0)+\ks g(\bks)+k\chi'(0))\right)'\right|\notag\\\leq&\left|\left(\left(\frac{k^2}{\kks}\right)'\frac{1}{\kks}(\chi(\bk)-\chi(0)-k\chi'(0))\right)'\right|\notag\\&+\left|\left(\left(\frac{k^2}{\kks}\right)'\frac{1}{\kks}(\ks g(\bks)+k\chi'(0))\right)'\right|\notag\\=:&\left|D_3^1\right|+\left|D_3^2\right|,
\end{align}
with appropriate bounded $g(\bks):$ By Taylors formula and since $|\nabla_{\bk}\chi(\bk)|\leq 3C$ we get 
\begin{align}\label{i3o}
|g(\bks)|&\leq 3C.
\end{align}
$D_3^2$ in (\ref{i3n}) can be treated analogous to $D_2$ since the derivative has at most five zeros and 
\begin{align}\label{i3p}
\left|\left(\frac{k^2}{\kks}\right)'\frac{1}{\kks}(\ks g(\bks)+k\chi'(0))\right|\leq&40C.
\end{align}
To get (\ref{i3p}) we use again estimates like (\ref{substi}) with $k\geq 2\ks.$
Now we estimate $D_3^1$ in (\ref{i3n}). Since the integration area is bounded it suffices to show that $D_3^1$ is uniformly bounded:
\begin{align}\label{i3q}
\left|D_3^1\right|\leq &\left|\left(\frac{k^2}{\kks}\right)''\frac{1}{\kks}(\chi(\bk)-\chi(0)-k\chi'(0))\right|\notag\\&+\left|\left(\frac{k^2}{\kks}\right)'\left(\frac{1}{\kks}\right)'(\chi(\bk)-\chi(0)-k\chi'(0))\right|\notag\\&+\left|\left(\frac{k^2}{\kks}\right)'\frac{1}{\kks}(\chi'(\bk)-\chi'(0))\right|.
\end{align}
Using Taylors formula we can linearize the $\chi(\bk)$-terms and get: ($0<\xi,\zeta<1$)
\begin{align}\label{i3r}
\left|D_3^1\right|\leq &\left|\left(\frac{k^2}{\kks}\right)''\frac{1}{\kks}k^2\chi''(\xi \bk)\right|+\left|\left(\frac{k^2}{\kks}\right)'\left(\frac{1}{\kks}\right)'k^2\chi''(\xi\bk)\right|\notag\\&+\left|\left(\frac{k^2}{\kks}\right)'\frac{1}{\kks}k\chi''(\zeta \bk)\right|.
\end{align}
Using $|\chi''(\bk)|\leq C$ and (\ref{substi}) (again we use also similar estimates with $k\geq 2\ks$) one gets:
\begin{equation}
\left|D_3^1\right|\leq 120C.
\end{equation}
Hence, $I_3$ is of order $t^{-2}$ uniform in $\bks.$ It follows Lemma \ref{lemstatphas}.
\epro

\end{document}